\documentclass[aip,jcp,reprint,noshowkeys,superscriptaddress]{revtex4-1}
\usepackage{graphicx,dcolumn,bm,xcolor,microtype,multirow,amsmath,amssymb,amsfonts,physics,mhchem}

\usepackage[utf8]{inputenc}
\usepackage[T1]{fontenc}
\usepackage{txfonts}
\usepackage{mathrsfs}

\usepackage[
	colorlinks=true,
    citecolor=blue,
    breaklinks=true
	]{hyperref}
\usepackage[normalem]{ulem}
\newcommand{\titou}[1]{\textcolor{black}{#1}}

\newcommand{\manurev}[1]{\textcolor{black}{#1}}

\newcommand{\tabc}[1]{\multicolumn{1}{c}{#1}}
\newcommand{\la}{\lambda}
\newcommand{\si}{\sigma}
\newcommand{\ie}{\textit{i.e.}}

\newcommand{\hH}{\Hat{H}}
\newcommand{\hh}{\Hat{h}}

\newcommand{\vne}{v_\text{ne}}
\newcommand{\hWee}{\Hat{W}_\text{ee}}
\newcommand{\WHF}{W_\text{HF}}

\newcommand{\eps}{\epsilon}
\newcommand{\e}[2]{\eps_\text{#1}^{#2}}
\newcommand{\E}[2]{E_\text{#1}^{#2}}

\newcommand{\be}[2]{\overline{\eps}_\text{#1}^{#2}}

\newcommand{\n}[2]{n_{#1}^{#2}}

\newcommand{\DD}[2]{\Delta_\text{#1}^{#2}}

\newcommand{\br}[1]{\boldsymbol{r}_{#1}}

\newcommand{\bw}{{\boldsymbol{w}}}
\newcommand{\bG}{\boldsymbol{G}}
\newcommand{\bS}{\boldsymbol{S}}
\newcommand{\bGam}[1]{\boldsymbol{\Gamma}^{#1}}
\newcommand{\bgam}[1]{\boldsymbol{\gamma}^{#1}}
\newcommand{\opGam}[1]{\hat{\Gamma}^{#1}}
\newcommand{\bh}{\boldsymbol{h}}
\newcommand{\bF}[1]{\boldsymbol{F}^{#1}}
\newcommand{\Ex}[2]{\Omega_\text{#1}^{#2}}

\newcommand{\ew}[1]{w_{#1}}
\newcommand{\eG}[1]{G_{#1}}
\newcommand{\eS}[1]{S_{#1}}
\newcommand{\eGam}[2]{\Gamma_{#1}^{#2}}

\newcommand{\eh}[2]{h_{#1}^{#2}}
\newcommand{\eF}[2]{F_{#1}^{#2}}
\newcommand{\ERI}[2]{(#1|#2)}
\newcommand{\dbERI}[2]{(#1||#2)}

\newcommand{\nEl}{N}
\newcommand{\nBas}{K}

\newcommand{\Det}[1]{\Phi^{#1}}
\newcommand{\MO}[2]{\phi_{#1}^{#2}}
\newcommand{\SO}[2]{\varphi_{#1}^{#2}}
\newcommand{\cMO}[2]{c_{#1}^{#2}}
\newcommand{\AO}[1]{\chi_{#1}}

\newcommand{\SI}{\textcolor{blue}{supplementary material}}

\newcommand{\LCPQ}{Laboratoire de Chimie et Physique Quantiques (UMR 5626), Universit\'e de Toulouse, CNRS, UPS, France}
\newcommand{\LCQ}{Laboratoire de Chimie Quantique, Institut de Chimie, CNRS, Universit\'e de Strasbourg, Strasbourg, France}

\newcommand{\beq}{\begin{equation}}
\newcommand{\eeq}{\end{equation}}
\newcommand{\bxi}{\boldsymbol{\xi}}

\DeclareMathOperator*{\argmin}{arg\,min}

\begin{document}	

\title{A weight-dependent local correlation density-functional approximation for ensembles}

\author{Pierre-Fran\c{c}ois Loos}
\email{loos@irsamc.ups-tlse.fr}
\affiliation{\LCPQ}
\author{Emmanuel Fromager}
\email{fromagere@unistra.fr}
\affiliation{\LCQ}

\begin{abstract}
We report a local, weight-dependent correlation density-functional approximation that incorporates information about both ground and excited states in the context of density-functional theory for ensembles (eDFT). 
This density-functional approximation for ensembles is specially
designed for the computation of single and double excitations within
Gross--Oliveira--Kohn (GOK) DFT (\textit{i.e.}, eDFT for neutral
excitations), and can be seen as a natural extension of the ubiquitous local-density approximation in the context of ensembles.
The resulting density-functional approximation, based on both finite and infinite uniform electron gas models, automatically incorporates the infamous derivative discontinuity contributions to the excitation energies through its explicit ensemble weight dependence. 
Its accuracy is illustrated by computing single and double excitations in one-dimensional many-electron systems in the weak, intermediate and strong correlation regimes.
Although the present weight-dependent functional has been specifically
designed for one-dimensional systems, the methodology proposed here is
general, \ie, directly applicable to the construction of weight-dependent functionals for realistic three-dimensional systems, such as molecules and solids.
\end{abstract}

\maketitle

\section{Introduction}
\label{sec:intro}
Over the last two decades, density-functional theory (DFT)
\cite{Hohenberg_1964,Kohn_1965,ParrBook} has become the method of choice for
modeling the electronic structure of large molecular systems and
materials.
The main reason is that, within DFT, the quantum contributions to the
electronic repulsion energy --- the so-called exchange-correlation (xc)
energy --- is rewritten as a functional of the electron density $\n{}{} \equiv \n{}{}(\br{})$, the latter being a much simpler quantity than the many-electron wave function. 
The complexity of the many-body problem is then transferred to the xc
density functional. 
Despite its success, the standard Kohn-Sham (KS) formulation of DFT \cite{Kohn_1965} (KS-DFT) suffers, in practice, from various deficiencies. \cite{Woodcock_2002, Tozer_2003,Tozer_1999,Dreuw_2003,Sobolewski_2003,Dreuw_2004,Tozer_1998,Tozer_2000,Casida_1998,Casida_2000,Tapavicza_2008,Levine_2006}
The description of strongly multiconfigurational ground states (often
referred to as ``strong correlation problem'') still remains a
challenge. \cite{Gori-Giorgi_2010,Fromager_2015,Gagliardi_2017}
Another issue, which is partly connected to the previous one, is the
description of low-lying quasi-degenerate states. 

The standard approach for modeling excited states in a DFT framework is
linear-response time-dependent DFT (TDDFT). \cite{Runge_1984,Casida,Casida_2012}
In this case, the electronic spectrum relies on the (unperturbed) pure-ground-state KS picture, which may break down when electron correlation is strong. 
Moreover, in exact TDDFT, the xc energy is in fact an xc {\it action} \cite{Vignale_2008} which is a
functional of the time-dependent density $\n{}{} \equiv \n{}{}(\br,t)$ and, as
such, it should incorporate memory effects. Standard implementations of TDDFT rely on 
the adiabatic approximation where these effects are neglected. \cite{Dreuw_2005} In other
words, the xc functional is assumed to be local in time. \cite{Casida,Casida_2012}
As a result, double electronic excitations (where two electrons are simultaneously promoted by a single photon) are completely absent from the TDDFT spectrum, thus reducing further the applicability of TDDFT. \cite{Maitra_2004,Cave_2004,Mazur_2009,Romaniello_2009a,Sangalli_2011,Mazur_2011,Huix-Rotllant_2011,Elliott_2011,Maitra_2012,Sundstrom_2014,Loos_2019}

When affordable (\ie, for relatively small molecules), time-independent
state-averaged wave function methods
\cite{Roos,Andersson_1990,Angeli_2001a,Angeli_2001b,Angeli_2002,Helgakerbook} can be employed to fix the various issues mentioned above. 
The basic idea is to describe a finite (canonical) ensemble of ground
and excited states altogether, \ie, with the same set of orbitals. 
Interestingly, a similar approach exists in DFT. Referred to as
Gross--Oliveira--Kohn (GOK) DFT, \cite{Gross_1988a,Gross_1988b,Oliveira_1988} it was proposed at the end of the 80's as a generalization
of Theophilou's DFT for equiensembles. \cite{Theophilou_1979}
In GOK-DFT, the ensemble xc energy is a functional of the
density {\it and} a  
function of the ensemble weights. Note that, unlike in conventional
Boltzmann ensembles, \cite{Pastorczak_2013} the ensemble weights (each state in the ensemble
is assigned a given and fixed weight) are allowed to vary
independently in a GOK ensemble.   
The weight dependence of the xc functional plays a crucial role in the
calculation of excitation energies.
\cite{Gross_1988b,Yang_2014,Deur_2017,Deur_2019,Senjean_2018,Senjean_2020} 
It actually accounts for the derivative discontinuity contribution to energy gaps. \cite{Levy_1995, Perdew_1983}

Even though GOK-DFT is in principle able to
describe near-degenerate situations and multiple-electron excitation
processes, it has not
been given much attention until quite recently. \cite{Franck_2014,Borgoo_2015,Kazaryan_2008,Gould_2013,Gould_2014,Filatov_2015,Filatov_2015b,Filatov_2015c,Gould_2017,Deur_2017,Gould_2018,Gould_2019,Sagredo_2018,Ayers_2018,Deur_2018,Deur_2019,Kraisler_2013, Kraisler_2014,Alam_2016,Alam_2017,Nagy_1998,Nagy_2001,Nagy_2005,Pastorczak_2013,Pastorczak_2014,Pribram-Jones_2014,Yang_2013a,Yang_2014,Yang_2017,Senjean_2015,Senjean_2016,Senjean_2018,Smith_2016}
One of the reason is the lack, not to say the absence, of reliable
density-functional approximations for ensembles. 
The most recent works dealing with this particular issue are still fundamental and
exploratory, as they rely either on simple (but nontrivial) model
systems
\cite{Carrascal_2015,Deur_2017,Deur_2018,Deur_2019,Senjean_2015,Senjean_2016,Senjean_2018,Sagredo_2018,Senjean_2020,Fromager_2020,Gould_2019}
or atoms. \cite{Yang_2014,Yang_2017,Gould_2019_insights}
Despite all these efforts, it is still unclear how weight dependencies
can be incorporated into density-functional approximations. This problem is actually central not
only in GOK-DFT but also in conventional (ground-state) DFT as the infamous derivative
discontinuity problem that occurs when crossing an integral number of
electrons can be recast into a weight-dependent ensemble
one. \cite{Senjean_2018,Senjean_2020}      

The present work is an attempt to address the ensemble weight dependence problem
in GOK-DFT,  
with the ambition to turn the theory, in the forthcoming future, into a
(low-cost) practical computational method for modeling excited states in molecules and extended systems.
Starting from the ubiquitous local-density approximation (LDA), we
design a weight-dependent ensemble correction based on a finite uniform
electron gas from which density-functional excitation energies can be
extracted. The present density-functional approximation for ensembles, which can be seen as a natural
extension of the LDA, will be referred to as eLDA in the remaining of this paper.
As a proof of concept, we apply this general strategy to
ensemble correlation energies (that we combine with 
ensemble exact exchange energies) in the particular case of
\emph{strict} one-dimensional (1D)
spin-polarized systems. \cite{Loos_2012, Loos_2013a, Loos_2014a, Loos_2014b} 
In other words, the Coulomb interaction used in this work corresponds to
particles which are \emph{strictly} restricted to move within a 1D sub-space of three-dimensional space.
Despite their simplicity, 1D models are scrutinized as paradigms for quasi-1D materials \cite{Schulz_1993, Fogler_2005a} such as carbon nanotubes \cite{Bockrath_1999, Ishii_2003, Deshpande_2008} or nanowires. \cite{Meyer_2009, Deshpande_2010}
This description of 1D systems also has interesting connections with the exotic chemistry of ultra-high magnetic fields (such as those in white dwarf stars), where the electronic cloud is dramatically compressed perpendicular to the magnetic field. \cite{Schmelcher_1990, Lange_2012, Schmelcher_2012}
In these extreme conditions, where magnetic effects compete with Coulombic forces, entirely new bonding paradigms emerge. \cite{Schmelcher_1990, Schmelcher_1997, Tellgren_2008, Tellgren_2009, Lange_2012, Schmelcher_2012, Boblest_2014, Stopkowicz_2015}

The paper is organized as follows. 
Exact and approximate formulations of GOK-DFT are discussed in Sec.~\ref{sec:eDFT}, 
with a particular emphasis on the extraction of individual energy levels.
In Sec.~\ref{sec:eDFA}, we detail the construction of the
weight-dependent local correlation functional specially designed for the
computation of single and double excitations within GOK-DFT.
Computational details needed to reproduce the results of the present work are reported in Sec.~\ref{sec:comp_details}.
In Sec.~\ref{sec:res}, we illustrate the accuracy of the present eLDA functional by computing single and double excitations in 1D many-electron systems in the weak, intermediate and strong correlation regimes.
Finally, we draw our conclusions in Sec.~\ref{sec:conclusion}.
Atomic units are used throughout.

\section{Theory}
\label{sec:eDFT}

\subsection{GOK-DFT}\label{subsec:gokdft}

In this section we give a brief review of GOK-DFT and discuss the
extraction of individual energy levels \cite{Deur_2019,Fromager_2020} with a particular focus on exact
individual exchange energies. 
Let us start by introducing the GOK ensemble energy \cite{Gross_1988a}
\beq\label{eq:exact_GOK_ens_ener}
	\E{}{\bw}=\sum_{K \geq 0} \ew{K} \E{}{(K)},
\eeq
where the $K$th energy level $\E{}{(K)}$ [$K=0$ refers to the ground state] is the eigenvalue of the electronic Hamiltonian $\hH = \hh + \hWee$, where 
\beq
	\hh = \sum_{i=1}^\nEl \qty[ -\frac{1}{2} \nabla_{i}^2 + \vne(\br{i}) ]
\eeq
is the one-electron operator describing kinetic and nuclear attraction energies, and $\hat{W}_{\rm ee}$ is the electron repulsion operator. 
The (positive) ensemble weights $\ew{K}$ decrease with increasing index $K$. 
They are normalized, \ie,
\beq\label{eq:weight_norm_cond}
	\ew{0} = 1 - \sum_{K>0} \ew{K},
\eeq
so that only the weights $\bw \equiv \qty( \ew{1}, \ew{2}, \ldots, \ew{K}, \ldots )$ assigned to the excited states can vary independently.  
For simplicity we will assume in the following that the energies are not degenerate. 
Note that the theory can be extended to multiplets simply by assigning the same ensemble weight to all degenerate states.\cite{Gross_1988b}
In the KS formulation of GOK-DFT, {which is simply referred to as
KS ensemble DFT (KS-eDFT) in the following}, the ensemble energy is determined variationally as follows:\cite{Gross_1988b}
\beq\label{eq:var_ener_gokdft}
	\E{}{\bw}
 	= \min_{\opGam{\bw}}
	\qty{ 
	\Tr[\opGam{\bw} \hh] + \E{Hx}{\bw} \qty[\n{\opGam{\bw}}{}] + \E{c}{\bw} \qty[\n{\opGam{\bw}}{}]
	},
\eeq
where $\Tr$ denotes the trace and the trial ensemble density matrix operator reads
\beq
	\opGam{\bw}=\sum_{K \geq 0} \ew{K} \dyad*{\Det{(K)}}.
\eeq
The KS determinants [or configuration state functions~\cite{Gould_2017}]
$\Det{(K)}$ are all constructed from the same set of ensemble KS
orbitals that are variationally optimized.
The trial ensemble density in Eq.~\eqref{eq:var_ener_gokdft} is simply
the weighted sum of the individual KS densities, \ie,
\beq\label{eq:KS_ens_density}
	\n{\opGam{\bw}}{}(\br{}) = \sum_{K\geq 0} \ew{K} \n{\Det{(K)}}{}(\br{}).
\eeq
As readily seen from Eq.~\eqref{eq:var_ener_gokdft}, both Hartree-exchange (Hx) and correlation (c) energies are described with density functionals that are \textit{weight dependent}. 
We focus in the following on the (exact) Hx part, which is defined as~\cite{Gould_2017}
\beq\label{eq:exact_ens_Hx}
	\E{Hx}{\bw}[\n{}{}]=\sum_{K \geq 0} \ew{K} \mel*{\Det{(K),\bw}[\n{}{}]}{\hWee}{\Det{(K),\bw}[\n{}{}]},
\eeq
where the KS wavefunctions fulfill the ensemble density constraint
\beq
	\sum_{K\geq 0} \ew{K} \n{\Det{(K),\bw}[\n{}{}]}{}(\br{}) = \n{}{}(\br{}).
\eeq 
The (approximate) description of the correlation part is discussed in
Sec.~\ref{sec:eDFA}.

In practice, the ensemble energy is not the most interesting quantity, and one is more concerned with excitation energies or individual energy levels (for geometry optimizations, for example). 
As pointed out recently in Ref.~\onlinecite{Deur_2019}, the latter can be extracted
exactly from a single ensemble calculation as follows:
\beq\label{eq:indiv_ener_from_ens}
	\E{}{(I)} = \E{}{\bw} + \sum_{K>0} \qty(\delta_{IK} - \ew{K} )
\pdv{\E{}{\bw}}{\ew{K}},
\eeq
where, according to the normalization condition of Eq.~\eqref{eq:weight_norm_cond}, 
\beq
\pdv{\E{}{\bw}}{\ew{K}}= \E{}{(K)} -
\E{}{(0)}\equiv\Ex{}{(K)}
\eeq
corresponds to the $K$th excitation energy.
According to the {\it variational} ensemble energy expression of
Eq.~\eqref{eq:var_ener_gokdft}, the derivative with respect to $\ew{K}$
can be evaluated from the minimizing weight-dependent KS wavefunctions
$\Det{(K)} \equiv \Det{(K),\bw}$ as follows:
\beq\label{eq:deriv_Ew_wk}
\begin{split}
	\pdv{\E{}{\bw}}{\ew{K}} 
	& = \mel*{\Det{(K)}}{\hh}{\Det{(K)}}-\mel*{\Det{(0)}}{\hh}{\Det{(0)}}
	\\
	& + \Bigg\{\int \fdv{\E{Hx}{\bw}[\n{}{}]}{\n{}{}(\br{})} \qty[ \n{\Det{(K)}}{}(\br{}) - \n{\Det{(0)}}{}(\br{}) ] d\br{}
	+ \pdv{\E{Hx}{\bw} [\n{}{}]}{\ew{K}}
	\\
	& + \int \fdv{\E{c}{\bw}[n]}{\n{}{}(\br{})} \qty[ \n{\Det{(K)}}{}(\br{}) - \n{\Det{(0)}}{}(\br{}) ] d\br{} 
	+ \pdv{\E{c}{\bw}[n]}{\ew{K}}
	\Bigg\}_{\n{}{} = \n{\opGam{\bw}}{}}.
\end{split}
\eeq
The Hx contribution from Eq.~\eqref{eq:deriv_Ew_wk} can be recast as
\beq\label{eq:_deriv_wk_Hx}
	\left.
	\pdv{}{\xi_K} \qty(\E{Hx}{\bxi} [\n{}{\bxi,\bxi}]
	- \E{Hx}{\bw}[\n{}{\bw,\bxi}] )
	\right|_{\bxi=\bw},
\eeq
where $\bxi \equiv (\xi_1,\xi_2,\ldots,\xi_K,\ldots)$ and the
auxiliary double-weight ensemble density reads
\beq
	\n{}{\bw,\bxi}(\br{}) = \sum_{K\geq 0} \ew{K} \n{\Det{(K),\bxi}}{}(\br{}).
\eeq 
Since, for given ensemble weights $\bw$ and $\bxi$, the ensemble
densities $\n{}{\bxi,\bxi}$ and $\n{}{\bw,\bxi}$ are obtained from the \textit{same} KS potential (which is unique up to a constant), it comes
from the exact expression in Eq.~\eqref{eq:exact_ens_Hx} that 
\beq
	\E{Hx}{\bxi}[\n{}{\bxi,\bxi}] = \sum_{K \geq 0} \xi_K \mel*{\Det{(K),\bxi}}{\hWee}{\Det{(K),\bxi}},
\eeq  
and
\beq
	\E{Hx}{\bw}[\n{}{\bw,\bxi}] = \sum_{K \geq 0} \ew{K} \mel*{\Det{(K),\bxi}}{\hWee}{\Det{(K),\bxi}}.
\eeq  
This yields, according to Eqs.~\eqref{eq:deriv_Ew_wk} and \eqref{eq:_deriv_wk_Hx}, the simplified expression
\beq\label{eq:deriv_Ew_wk_simplified}
\begin{split}
	\pdv{\E{}{\bw}}{\ew{K}} 
	& = \mel*{\Det{(K)}}{\hH}{\Det{(K)}} 
	- \mel*{\Det{(0)}}{\hH}{\Det{(0)}}
	\\
	& + \qty{
	\int \fdv{\E{c}{\bw}[\n{}{}]}{\n{}{}({\br{}})} 
	\qty[ \n{\Det{(K)}}{}(\br{}) - \n{\Det{(0)}}{}(\br{}) ]
	+
	\pdv{\E{c}{\bw} [\n{}{}]}{\ew{K}} 
	}_{\n{}{} = \n{\opGam{\bw}}{}} d\br{}.
\end{split}
\eeq
Since, according to Eqs.~\eqref{eq:var_ener_gokdft} and \eqref{eq:exact_ens_Hx}, the ensemble energy can be evaluated as 
\beq
	\E{}{\bw} = \sum_{K \geq 0} \ew{K} \mel*{\Det{(K)}}{\hH}{\Det{(K)}} + \E{c}{\bw}[\n{\opGam{\bw}}{}],
\eeq 
with $\Det{(K)} = \Det{(K),\bw}$ [note that, when the minimum is reached in Eq.~\eqref{eq:var_ener_gokdft}, $\n{\opGam{\bw}}{} = \n{}{\bw,\bw}$], 
we finally recover from Eqs.~\eqref{eq:KS_ens_density} and
\eqref{eq:indiv_ener_from_ens} the {\it exact} expression of Ref.~\onlinecite{Fromager_2020} for the $I$th energy level:
\beq\label{eq:exact_ener_level_dets}
\begin{split}
	\E{}{(I)} 
	& = \mel*{\Det{(I)}}{\hH}{\Det{(I)}} + \E{c}{{\bw}}[\n{\opGam{\bw}}{}]
	\\
	& + \int \fdv{\E{c}{\bw}[\n{\opGam{\bw}}{}]}{\n{}{}(\br{})}
	\qty[ \n{\Det{(I)}}{}(\br{}) - \n{\opGam{\bw}}{}(\br{}) ] d\br{}
	\\
	&+
	\sum_{K>0} \qty(\delta_{IK} - \ew{K} )
	\left.
	\pdv{\E{c}{\bw}[\n{}{}]}{\ew{K}}
	\right|_{\n{}{} = \n{\opGam{\bw}}{}}.
\end{split}
\eeq
Note that, when $\bw=0$, the ensemble correlation functional reduces to the
conventional (ground-state) correlation functional $E_{\rm c}[n]$. As a
result, the regular KS-DFT expression is recovered from
Eq.~\eqref{eq:exact_ener_level_dets} for the ground-state energy:
\beq
\E{}{(0)}=\mel*{\Det{(0)}}{\hH}{\Det{(0)}} +
\E{c}{}[\n{\Det{(0)}}{}],
\eeq
or, equivalently,
\beq\label{eq:gs_ener_level_gs_lim}
\E{}{(0)}=\mel*{\Det{(0)}}{\hat{H}[\n{\Det{(0)}}{}]}{\Det{(0)}} 
,
\eeq
where the density-functional Hamiltonian reads
\beq\label{eq:dens_func_Hamilt}
\hat{H}[n]=\hH+
\sum^N_{i=1}\left(\fdv{\E{c}{}[n]}{\n{}{}(\br{i})}
+C_{\rm c}[n]
\right),
\eeq
and 
\beq\label{eq:corr_LZ_shift}
C_{\rm c}[n]=\dfrac{\E{c}{}[n]
        -\int
\fdv{\E{c}{}[n]}{\n{}{}(\br{})}n(\br{})d\br{}}{\int n(\br{})d\br{}}
\eeq
is the correlation component of
Levy--Zahariev's constant shift in potential.\cite{Levy_2014}
Similarly, the excited-state ($I>0$) energy level expressions
can be recast as follows: 
\beq\label{eq:excited_ener_level_gs_lim}
	\E{}{(I)} 
	 = \mel*{\Det{(I)}}{\hat{H}[\n{\Det{(0)}}{}]}{\Det{(I)}} 
      +
      \left. 
	\pdv{\E{c}{\bw}[\n{\Det{(0)}}{}]}{\ew{I}}
	\right|_{\bw=0}.
\eeq 
As readily seen from Eqs.~\eqref{eq:dens_func_Hamilt} and
\eqref{eq:corr_LZ_shift}, introducing any constant shift $\delta
\E{c}{}[\n{\Det{(0)}}{}]/\delta n({\bf r})\rightarrow \delta
\E{c}{}[\n{\Det{(0)}}{}]/\delta n({\bf r})+C$ into the correlation
potential leaves the density-functional Hamiltonian $\hat{H}[n]$ (and
therefore the individual energy levels) unchanged. As a result, in
this context,
the correlation derivative discontinuities induced by the 
excitation process~\cite{Levy_1995} will be fully described by the 
correlation ensemble derivatives [second term on the right-hand side of
Eq.~\eqref{eq:excited_ener_level_gs_lim}].    

\subsection{One-electron reduced density matrix formulation}
For implementation purposes, we will use in the rest of this work 
(one-electron reduced) density matrices 
as basic variables, rather than Slater determinants. 
As the theory is applied later on to \textit{spin-polarized}
systems, we drop spin indices in the density matrices, for convenience.
If we expand the
ensemble KS orbitals (from which the determinants are constructed) in an atomic orbital (AO) basis,
\beq
	\MO{p}{}(\br{}) =  \sum_{\mu} \cMO{\mu p}{} \AO{\mu}(\br{}),
\eeq
then the density matrix of the   
determinant $\Det{(K)}$ can be expressed as follows in the AO basis:
\beq
	\bGam{(K)} \equiv \eGam{\mu\nu}{(K)} = \sum_{\SO{p}{} \in (K)} \cMO{\mu p}{} \cMO{\nu p}{},
\eeq
where the summation runs over the orbitals that are occupied in $\Det{(K)}$. 
The electron density of the $K$th KS determinant can then be evaluated
as follows:
\beq
	\n{\bGam{(K)}}{}(\br{}) = \sum_{\mu\nu} \AO{\mu}(\br{}) \eGam{\mu\nu}{(K)} \AO{\nu}(\br{}),
\eeq 
while the ensemble density matrix
and the ensemble density read
\beq\label{eq:ens1RDM}
	\bGam{\bw} 
	= \sum_{K\geq 0} \ew{K} \bGam{(K)}
	\equiv \eGam{\mu\nu}{\bw}
	= \sum_{K\geq 0} \ew{K} \eGam{\mu\nu}{(K)},
\eeq
and
\beq\label{eq:ens_dens_from_ens_1RDM}
	\n{\bGam{\bw}}{}(\br{}) = \sum_{\mu\nu} \AO{\mu}(\br{}) \eGam{\mu\nu}{\bw} \AO{\nu}(\br{}),
\eeq
respectively. 
The exact individual energy expression in Eq.~\eqref{eq:exact_ener_level_dets} can then be rewritten as
\beq\label{eq:exact_ind_ener_rdm}
\begin{split}
	\E{}{(I)} 
	& =\Tr[\bGam{(I)} \bh]
	+ \frac{1}{2} \Tr[\bGam{(I)} \bG \bGam{(I)}]
	+ \E{c}{{\bw}}[\n{\bGam{\bw}}{}]
	\\
	& + \int \fdv{\E{c}{\bw}[\n{\bGam{\bw}}{}]}{\n{}{}(\br{})} 
	\qty[ \n{\bGam{(I)}}{}(\br{}) - \n{\bGam{\bw}}{}(\br{}) ] d\br{} 
	\\
	& + \sum_{K>0} \qty(\delta_{IK} - \ew{K})
	\left. \pdv{\E{c}{\bw}[\n{}{}]}{\ew{K}}\right|_{\n{}{} = \n{\bGam{\bw}}{}}
	,
\end{split}
\eeq
where 
\beq
	\bh \equiv h_{\mu\nu} = \mel*{\AO{\mu}}{\hh}{\AO{\nu}}	
\eeq
denotes the matrix of the one-electron integrals.
The exact individual Hx energies are obtained from the following trace formula
\beq
	\Tr[\bGam{(K)} \bG \bGam{(L)}]
	= \sum_{\mu\nu\la\si} \eGam{\mu\nu}{(K)} \eG{\mu\nu\la\si} \eGam{\la\si}{(L)},
\eeq
where the antisymmetrized two-electron integrals read
\beq
	\bG 
	\equiv G_{\mu\nu\la\si} 
	= \dbERI{\mu\nu}{\la\si} 
	= \ERI{\mu\nu}{\la\si} - \ERI{\mu\si}{\la\nu},
\eeq
with
\beq
	\ERI{\mu\nu}{\la\si} = \iint \frac{\AO{\mu}(\br{1}) \AO{\nu}(\br{1}) \AO{\la}(\br{2}) \AO{\si}(\br{2})}{\abs{\br{1} - \br{2}}} d\br{1} d\br{2}.
\eeq

\subsection{Approximations}\label{subsec:approx}

In the following, GOK-DFT will be applied
to 1D
spin-polarized systems where 
Hartree and exchange energies cannot be separated. 
For that reason, we will substitute the Hartree--Fock (HF) density-matrix-functional interaction energy,
\beq\label{eq:eHF-dens_mat_func}
	\WHF[\bGam{}] = \frac{1}{2} \Tr[\bGam{} \bG \bGam{}], 
\eeq
for the Hx density-functional energy in the variational energy
expression of Eq.~\eqref{eq:var_ener_gokdft}, thus leading to the
following approximation:  
\beq\label{eq:min_with_HF_ener_fun}
	\bGam{\bw} 
	\rightarrow \argmin_{\bgam{\bw}} 
	\qty{ 
	\Tr[\bgam{\bw} \bh ] + \WHF[ \bgam{\bw}] + \E{c}{\bw}[\n{\bgam{\bw}}{}]
	}.
\eeq
The minimizing ensemble density matrix in Eq.~\eqref{eq:min_with_HF_ener_fun} fulfills the following
stationarity condition
\beq\label{eq:commut_F_AO}
	\bF{\bw} \bGam{\bw} \bS = \bS \bGam{\bw} \bF{\bw},
\eeq
where $\bS \equiv \eS{\mu\nu} = \braket*{\AO{\mu}}{\AO{\nu}}$ is the
overlap matrix and the ensemble Fock-like matrix reads
\beq
	\bF{\bw} \equiv \eF{\mu\nu}{\bw} = \eh{\mu\nu}{\bw} +
\sum_{\la\si} \eG{\mu\nu\la\si} \eGam{\la\si}{\bw},
\eeq 
with
\beq
	\eh{\mu\nu}{\bw} 
	= \eh{\mu\nu}{} + \int \AO{\mu}(\br{}) \fdv{\E{c}{\bw}[\n{\bGam{\bw}}{}]}{\n{}{}(\br{})} \AO{\nu}(\br{}) d\br{}.
\eeq

Note that, within the approximation of Eq.~\eqref{eq:min_with_HF_ener_fun}, the ensemble density matrix is
optimized with a non-local exchange potential rather than a
density-functional local one, as expected from
Eq.~\eqref{eq:var_ener_gokdft}. This procedure is actually general, \ie,
applicable to not-necessarily spin-polarized and real (higher-dimensional) systems. 
As readily seen from Eq.~\eqref{eq:eHF-dens_mat_func}, inserting the
ensemble density matrix into the HF interaction energy functional
introduces unphysical \textit{ghost-interaction} errors \cite{Gidopoulos_2002, Pastorczak_2014, Alam_2016, Alam_2017, Gould_2017}
as well as \textit{curvature}:\cite{Alam_2016,Alam_2017}
\beq\label{eq:WHF}
\begin{split}
	\WHF[\bGam{\bw}] 
	& = \frac{1}{2} \sum_{K\geq 0} \ew{K}^2 \Tr[\bGam{(K)} \bG \bGam{(K)}]
	\\
	& + \sum_{L>K\geq 0} \ew{K} \ew{L}\Tr[\bGam{(K)} \bG \bGam{(L)}].
\end{split}
\eeq
The ensemble energy is of course expected to vary linearly with the ensemble
weights [see Eq.~\eqref{eq:exact_GOK_ens_ener}].
The explicit linear weight dependence of the ensemble Hx energy is actually restored when evaluating the individual energy
levels on the basis of Eq.~\eqref{eq:exact_ind_ener_rdm}.

Turning to the density-functional ensemble correlation energy, the
following ensemble local-density approximation (eLDA) will be employed  
\beq\label{eq:eLDA_corr_fun}
	\E{c}{\bw}[\n{}{}]\approx \int \n{}{}(\br{}) \e{c}{\bw}(\n{}{}(\br{})) d\br{},
\eeq
where the \manurev{\textit{weight-dependent}} ensemble correlation
energy per particle \manurev{will have the general
expression} 
\beq\label{eq:decomp_ens_correner_per_part}
\e{c}{\bw}(\n{}{})=\sum_{K\geq 0}w_K\be{c}{(K)}(\n{}{}).
\eeq
\manurev{Note that, at this level of approximation, which is expected to
be exact for any \textit{uniform}
system, the 
density-functional correlation components $\be{c}{(K)}(\n{}{})$ are
weight-\textit{independent}, unlike in the exact theory. \cite{Fromager_2020}
As discussed further in Sec.~\ref{sec:eDFA}, these components can be
extracted from a
finite uniform electron gas model for which density-functional correlation excitation
energies can be computed. 
}\titou{Note also that, here, only the correlation part of the 
energy will be treated at the
DFT level while we rely on HF for the exchange part.
This is different from the usual context where both exchange and
correlation are treated at the LDA level which provides key error compensation features. 
As shown in Sec.~\ref{sec:res}, moving from the pure
ground-state picture to an equiensemble one can actually improve
the ground-state energy significantly within such a scheme, thus
highlighting a major difference between conventional and GOK DFT
calculations.}

The resulting KS-eLDA ensemble energy obtained via Eq.~\eqref{eq:min_with_HF_ener_fun}
reads 
\beq\label{eq:Ew-GIC-eLDA}
\E{eLDA}{\bw}=\Tr[\bGam{\bw}\bh] + \WHF[\bGam{\bw}] +\int
\e{c}{\bw}(\n{\bGam{\bw}}{}(\br{})) \n{\bGam{\bw}}{}(\br{}) d\br{}.
\eeq
Combining Eq.~\eqref{eq:exact_ind_ener_rdm} with
Eq.~\eqref{eq:eLDA_corr_fun} leads to our final expression of the
KS-eLDA energy levels 
\beq\label{eq:EI-eLDA}
\begin{split}
	\E{{eLDA}}{(I)} 
	= 
	\E{HF}{(I)} 
	+ \Xi_\text{c}^{(I)}
	+ \Upsilon_\text{c}^{(I)},
\end{split}
\eeq
where
\beq\label{eq:ind_HF-like_ener}
\E{HF}{(I)}=\Tr[\bGam{(I)} \bh] + \frac{1}{2} \Tr[\bGam{(I)} \bG \bGam{(I)}]
\eeq   
is the analog for ground and excited states (within an ensemble) of the HF energy, and
\begin{gather}
\begin{split}
\label{eq:Xic}
	\Xi_\text{c}^{(I)} 
	& = \int \e{c}{\bw}(\n{\bGam{\bw}}{}(\br{})) \n{\bGam{(I)}}{}(\br{}) d\br{}
	\\
	&
	+ \int \n{\bGam{\bw}}{}(\br{}) \qty[ \n{\bGam{(I)}}{}(\br{}) - \n{\bGam{\bw}}{}(\br{}) ]
	\left. \pdv{\e{c}{{\bw}}(\n{}{})}{\n{}{}} \right|_{\n{}{} =
\n{\bGam{\bw}}{}(\br{})} d\br{},
	\\
\end{split}
\\
\label{eq:Upsic}
	\Upsilon_\text{c}^{(I)} 
	= \int \sum_{K>0} \qty(\delta_{IK} - \ew{K} ) \n{\bGam{\bw}}{}(\br{})
	\left. \pdv{\e{c}{\bw}(\n{}{})}{\ew{K}} \right|_{\n{}{}=\n{\bGam{\bw}}{}(\br{})} d\br{}.
\end{gather}
\manurev{
One may naturally wonder about the physical content of the above correlation energy
expressions. It is in fact difficult to readily distinguish from 
Eqs.~\eqref{eq:Xic} and \eqref{eq:Upsic} purely (uncoupled) individual
contributions from mixed ones. For that purpose, we may 
consider a density regime which has a weak deviation from the uniform
one. In such a regime, where eLDA is a reasonable approximation, the
deviation of the individual densities from the ensemble one will be
small. As a result, 
we can} Taylor expand the density-functional
correlation contributions
around the $I$th KS state density
$\n{\bGam{(I)}}{}(\br{})$, \manurev{so that} the 
second term on the right-hand side
of Eq.~\eqref{eq:EI-eLDA} can be simplified as follows through first order in
$\n{\bGam{\bw}}{}(\br{})-\n{\bGam{(I)}}{}(\br{})$: 
\beq\label{eq:Taylor_exp_ind_corr_ener_eLDA}
	\Xi_\text{c}^{(I)}
	= \int \e{c}{\bw}(\n{\bGam{(I)}}{}(\br{})) \n{\bGam{(I)}}{}(\br{}) d\br{}
	+ \order{[\n{\bGam{\bw}}{}(\br{})-\n{\bGam{(I)}}{}(\br{})]^2}.
\eeq
Therefore, it can be identified as 
an individual-density-functional correlation energy where the density-functional
correlation energy per particle is approximated by the ensemble one for
all the states within the ensemble. \manurev{This perturbation expansion
is of course less relevant for (more realistic) systems that exhibit significant
deviations from the uniform
density regime. Nevertheless, it
gives more insight into the eLDA approximation and it becomes useful when
it comes to rationalize its performance, as illustrated in Sec. \ref{sec:res}.\\}  
Let us stress that, to the best of our knowledge, eLDA is the first
density-functional approximation that incorporates ensemble weight
dependencies explicitly, thus allowing for the description of derivative
discontinuities [see Eq.~\eqref{eq:excited_ener_level_gs_lim} and the
comment that follows] {\it via} the third term on the right-hand side
of Eq.~\eqref{eq:EI-eLDA}. According to the decomposition of
the ensemble
correlation energy per particle in Eq.
\eqref{eq:decomp_ens_correner_per_part}, the latter can be recast
\begin{equation}
\Upsilon_\text{c}^{(I)}
=\int 
\qty[\be{c}{(I)}(\n{\bGam{\bw}}{}(\br{}))
-
\e{c}{\bw}(\n{\bGam{\bw}}{}(\br{}))
] \n{\bGam{\bw}}{}(\br{})
 d\br{},
\end{equation}
thus leading to the following Taylor expansion through first order in
$\n{\bGam{\bw}}{}(\br{})-\n{\bGam{(I)}}{}(\br{})$: 
\beq\label{eq:Taylor_exp_DDisc_term}
\begin{split}
\Upsilon_\text{c}^{(I)}
&=
\int \qty[ \be{c}{(I)}(\n{\bGam{(I)}}{}(\br{})) - \e{c}{\bw}(\n{\bGam{(I)}}{}(\br{})) ] \n{\bGam{(I)}}{}(\br{}) d\br{}
\\
&+\int \Bigg[
\n{\bGam{(I)}}{}(\br{}) 
\left.\left(
\pdv{\be{c}{{(I)}}(\n{}{})}{\n{}{}} 
-
\pdv{\e{c}{{\bw}}(\n{}{})}{\n{}{}} 
\right)\right|_{\n{}{} =
\n{\bGam{(I)}}{}(\br{})}
\\
&+\be{c}{(I)}(\n{\bGam{(I)}}{}(\br{}))
-
\e{c}{\bw}(\n{\bGam{(I)}}{}(\br{}))\Bigg]
\qty[\n{\bGam{\bw}}{}(\br{})-\n{\bGam{(I)}}{}(\br{})]
d\br{}
\\
&
+ \order{[\n{\bGam{\bw}}{}(\br{})-\n{\bGam{(I)}}{}(\br{})]^2}.
\end{split}
\eeq
As readily seen from Eqs. \eqref{eq:Taylor_exp_ind_corr_ener_eLDA} and \eqref{eq:Taylor_exp_DDisc_term}, the 
role of the correlation ensemble derivative contribution $\Upsilon_\text{c}^{(I)}$ is, through zeroth order, to substitute the expected
individual correlation energy per particle for the ensemble one.

Let us finally mention that, while the weighted sum of the
individual KS-eLDA energy levels delivers a \textit{ghost-interaction-corrected} (GIC) version of
the KS-eLDA ensemble energy, \ie,
\beq\label{eq:Ew-eLDA}
\begin{split}
\E{GIC-eLDA}{\bw}&=\sum_{I\geq0}\ew{I}\E{{eLDA}}{(I)}
\\
&=
\E{eLDA}{\bw}
-\WHF[\bGam{\bw}]+\sum_{I\geq0}\ew{I}\WHF[ \bGam{(I)}],
\end{split}
\eeq
the excitation energies computed from the KS-eLDA individual energy level
expressions in Eq. \eqref{eq:EI-eLDA} can be simplified as follows: 
\beq\label{eq:Om-eLDA}
\begin{split}
	\Ex{eLDA}{(I)} 
	&= 
	\Ex{HF}{(I)}
\\
	&+ \int 
\qty[\e{c}{{\bw}}(\n{}{})+n\pdv{\e{c}{{\bw}}(\n{}{})}{\n{}{}}]	
_{\n{}{} =
\n{\bGam{\bw}}{}(\br{})}
\qty[ \n{\bGam{(I)}}{}(\br{}) - \n{\bGam{(0)}}{}(\br{}) ] d\br{} 
\\ & + \DD{c}{(I)},
\end{split}
\eeq
where the HF-like excitation energies, $\Ex{HF}{(I)} = \E{HF}{(I)} -
\E{HF}{(0)}$, are determined from a single set of ensemble KS orbitals and 
\beq\label{eq:DD-eLDA}
	\DD{c}{(I)}
	= \int \n{\bGam{\bw}}{}(\br{})
	\left. \pdv{\e{c}{\bw}(\n{}{})}{\ew{I}} \right|_{\n{}{}=\n{\bGam{\bw}}{}(\br{})} d\br{}
\eeq
is the eLDA correlation ensemble derivative contribution to the $I$th excitation energy.

\section{Density-functional approximations for ensembles}
\label{sec:eDFA}

\subsection{Paradigm}
\label{sec:paradigm}

Most of the standard local and semi-local density-functional approximations rely on the infinite uniform electron gas model (also known as jellium). \cite{ParrBook, Loos_2016}
One major drawback of the jellium paradigm, when it comes to develop density-functional approximations for ensembles, is that the ground and excited states are not easily accessible like in a molecule. \cite{Gill_2012, Loos_2012, Loos_2014a, Loos_2014b, Agboola_2015, Loos_2017a}
Moreover, because the infinite uniform electron gas model is a metal, it is gapless, which means that both the fundamental and optical gaps are zero.
From this point of view, using finite uniform electron gases, \cite{Loos_2011b,
Gill_2012} which have, like an atom, discrete energy levels and non-zero
gaps, can be seen as more relevant in this context. \cite{Loos_2014a, Loos_2014b, Loos_2017a}
However, an obvious drawback of using finite uniform electron gases is
that the resulting density-functional approximation for ensembles
will inexorably depend on the number of electrons in the finite uniform electron gas (see below).
Here, we propose to construct a weight-dependent LDA functional for the
calculation of excited states in 1D systems by combining finite uniform electron gases with the
usual infinite uniform electron gas paradigm.

As a finite uniform electron gas, we consider the ringium model in which electrons move on a perfect ring (\ie, a circle) but interact \textit{through} the ring. \cite{Loos_2012, Loos_2013a, Loos_2014b}
The most appealing feature of ringium regarding the development of
functionals in the context of GOK-DFT is the fact that both ground- and
excited-state densities are uniform, and therefore {\it equal}. 
As a result, the ensemble density will remain constant (and uniform) as the ensemble weights vary. 
This is a necessary condition for being able to model the 
correlation ensemble derivatives [last term
on the right-hand side of Eq.~\eqref{eq:exact_ener_level_dets}].
Moreover, it has been shown that, in the thermodynamic limit, the ringium model is equivalent to the ubiquitous infinite uniform electron gas paradigm. \cite{Loos_2013,Loos_2013a}
Let us stress that, in a finite uniform electron gas like ringium, the interacting and
noninteracting densities match individually for all the states within the
ensemble
(these densities are all equal to the uniform density), which means that
so-called density-driven correlation
effects~\cite{Gould_2019,Gould_2019_insights,Senjean_2020,Fromager_2020} are absent from the model.
Here, we will consider the most simple ringium system featuring electronic correlation effects, \ie, the two-electron ringium model.

The present weight-dependent density-functional approximation is specifically designed for the
calculation of excited-state energies within GOK-DFT.
To take into account both single and double excitations simultaneously, we consider a three-state ensemble including: 
(i) the ground state ($I=0$), (ii) the first singly-excited state ($I=1$), and (iii) the first doubly-excited state ($I=2$) of the (spin-polarized) two-electron ringium system.
To ensure the GOK variational principle, \cite{Gross_1988a} the
triensemble weights must fulfil the following conditions: \cite{Deur_2019}
$0 \le \ew{2} \le 1/3$ and $\ew{2} \le \ew{1} \le (1-\ew{2})/2$, where $\ew{1}$ and $\ew{2}$ are the weights associated with the singly- and doubly-excited states, respectively.
All these states have the same (uniform) density $\n{}{} = 2/(2\pi R)$, where $R$ is the radius of the ring on which the electrons are confined.
We refer the interested reader to Refs.~\onlinecite{Loos_2012, Loos_2013a, Loos_2014b} for more details about this paradigm.
Generalization to a larger number of states is straightforward and is left for future work.

\begin{table*}
	\caption{
	\label{tab:OG_func}
	Parameters of the weight-dependent correlation density-functional approximations defined in Eq.~\eqref{eq:ec}.}
		\begin{tabular}{lcddd}
			\hline\hline
			State					&	$I$		&	\tabc{$a_1^{(I)}$}	&	\tabc{$a_2^{(I)}$}	&	\tabc{$a_3^{(I)}$}	\\
			\hline
			Ground state			&	$0$		&	-0.0137078			&	0.0538982			&	0.0751740			\\
			Singly-excited state	&	$1$		&	-0.0238184			&	0.00413142			&	0.0568648			\\
			Doubly-excited state	&	$2$		&	-0.00935749			&	-0.0261936			&	0.0336645			\\
			\hline\hline
		\end{tabular}
\end{table*}

\subsection{Weight-dependent correlation functional}
\label{sec:Ec}

Based on highly-accurate calculations (see {\SI} for additional details), one can write down, for each state, an accurate analytical expression of the reduced (\ie, per electron) correlation energy \cite{Loos_2013a, Loos_2014a} via the following Pad\'e approximant
\begin{equation}
\label{eq:ec}
	\e{c}{(I)}(\n{}{}) = \frac{a_1^{(I)}\,\n{}{}}{\n{}{} + a_2^{(I)} \sqrt{\n{}{}} + a_3^{(I)}},
\end{equation}
where the $a_k^{(I)}$'s are state-specific fitting parameters provided in Table \ref{tab:OG_func}.
The value of $a_1^{(I)}$ is obtained via the exact high-density expansion of the correlation energy. \cite{Loos_2013a, Loos_2014a}
Equation \eqref{eq:ec} provides three state-specific correlation density-functional approximations based on a two-electron system.
Combining these, one can build the following three-state weight-dependent correlation density-functional approximation:
\begin{equation}
\label{eq:ecw}
 \Tilde{\epsilon}_{\rm c}^\bw(\n{}{})=  (1-\ew{1}-\ew{2}) \e{c}{(0)}(\n{}{}) + \ew{1} \e{c}{(1)}(\n{}{}) + \ew{2} \e{c}{(2)}(\n{}{}).
\end{equation}

\subsection{LDA-centered functional}

\begin{figure}
	\includegraphics[width=0.7\linewidth]{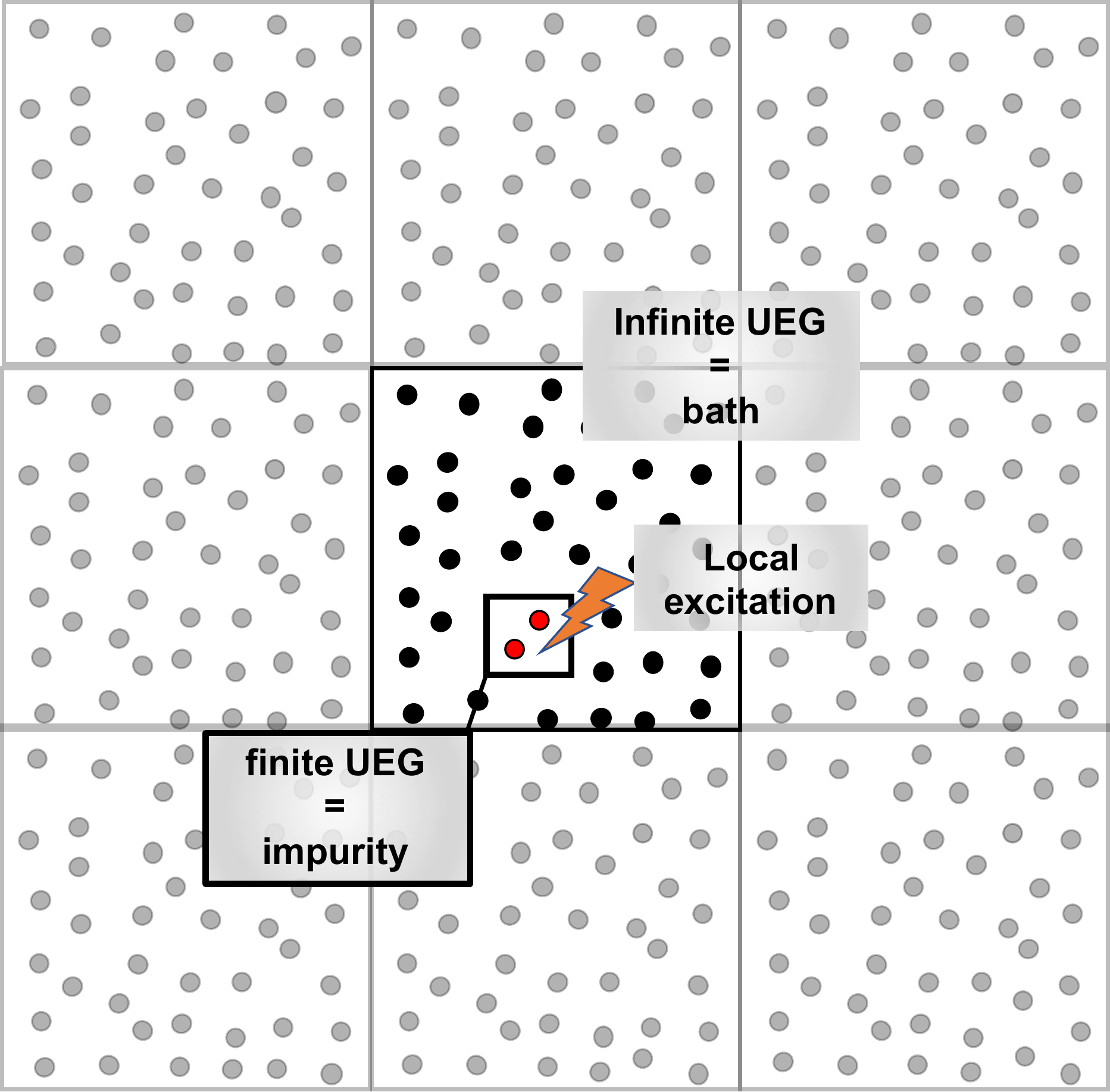}
	\caption{
	\label{fig:embedding}
	\titou{Schematic view of the ``embedding'' scheme: the two-electron finite uniform electron gas (the impurity) is embedded in the infinite uniform electron gas (the bath).
	The electronic excitation occurs locally, \ie, on the impurity.}
	}
\end{figure}

One of the main driving force behind the popularity of DFT is its ``universal'' nature, as xc density functionals can be applied to any electronic system.
Obviously, the two-electron-based density-functional approximation for ensemble defined in Eq.~\eqref{eq:ecw} does not have this feature as it does depend on the number of electrons constituting the finite uniform electron gas.
However, one can partially cure this dependency by applying a simple \titou{``embedding''} scheme \titou{(illustrated in Fig.~\ref{fig:embedding})} in which the two-electron finite uniform electron gas (the impurity) is embedded in the infinite uniform electron gas (the bath).
The weight-dependence of the correlation functional is then carried exclusively by the impurity [\ie, the functional defined in Eq.~\eqref{eq:ecw}], while the remaining correlation effects are provided by the bath (\ie, the usual LDA correlation functional).
Following this simple strategy, which can be further theoretically justified by the generalized adiabatic connection formalism for ensembles (GACE) originally derived by Franck and Fromager, \cite{Franck_2014} we propose to \emph{shift} the two-electron-based density-functional approximation for ensemble defined in Eq.~\eqref{eq:ecw} as follows:
\begin{equation}
\label{eq:becw}
	\Tilde{\epsilon}_{\rm c}^\bw(n)\rightarrow{\e{c}{\bw}(\n{}{})} =  (1-\ew{1}-\ew{2}) \be{c}{(0)}(\n{}{}) + \ew{1} \be{c}{(1)}(\n{}{}) + \ew{2} \be{c}{(2)}(\n{}{}),
\end{equation}
where
\begin{equation}
	\be{c}{(I)}(\n{}{}) = \e{c}{(I)}(\n{}{}) + \e{c}{\text{LDA}}(\n{}{}) - \e{c}{(0)}(\n{}{}).
\end{equation}
In the following, we will use the LDA correlation functional that has been specifically designed for 1D systems in
Ref.~\onlinecite{Loos_2013}:
\begin{equation}
\label{eq:LDA}
	\e{c}{\text{LDA}}(\n{}{}) 
	= a_1^\text{LDA} F\qty[1,\frac{3}{2},a_3^\text{LDA}, \frac{a_1^\text{LDA}(1-a_3^\text{LDA})}{a_2^\text{LDA}} {\n{}{}}^{-1}],
\end{equation}
where $F(a,b,c,x)$ is the Gauss hypergeometric function, \cite{NISTbook} and
\begin{subequations}
\begin{align}
	a_1^\text{LDA} & = - \frac{\pi^2}{360},
	\\
	a_2^\text{LDA} & = \frac{3}{4} - \frac{\ln{2\pi}}{2},
	\\ 
	a_3^\text{LDA} & = 2.408779.
\end{align}
\end{subequations}
Note that the strategy described in Eq.~\eqref{eq:becw} is general and
can be applied to real (higher-dimensional) systems. In order to make the
connection with the GACE formalism \cite{Franck_2014,Deur_2017} more explicit, one may  
recast Eq.~\eqref{eq:becw} as
\begin{equation}
\label{eq:eLDA}
\begin{split}
	{\e{c}{\bw}(\n{}{})}
	& =  \e{c}{\text{LDA}}(\n{}{}) 
	\\
	& + \ew{1} \qty[\e{c}{(1)}(\n{}{})-\e{c}{(0)}(\n{}{})] + \ew{2} \qty[\e{c}{(2)}(\n{}{})-\e{c}{(0)}(\n{}{})],
\end{split}
\end{equation}
or, equivalently,
\begin{equation}
\label{eq:eLDA_gace}
	{\e{c}{\bw}(\n{}{})}
	=  \e{c}{\text{LDA}}(\n{}{}) 
	+ \sum_{K>0}\int_0^{\ew{K}}
\qty[\e{c}{(K)}(\n{}{})-\e{c}{(0)}(\n{}{})]d\xi_K,
\end{equation}
where the $K$th correlation excitation energy (per electron) is integrated over the
ensemble weight $\xi_K$ at fixed (uniform) density $\n{}{}$. 
Equation \eqref{eq:eLDA_gace} nicely highlights the centrality of the
LDA in the present density-functional approximation for ensembles. 
In particular, ${\e{c}{(0,0)}(\n{}{})} = \e{c}{\text{LDA}}(\n{}{})$.
Consequently, in the following, we name this correlation functional ``eLDA'' as it is a natural extension of the LDA for ensembles.
Finally, we note that, by construction,
\begin{equation}
	{\pdv{\e{c}{\bw}(\n{}{})}{\ew{J}} = \e{c}{(J)}(\n{}{}) - \e{c}{(0)}(\n{}{}).}
\end{equation}

\section{Computational details}
\label{sec:comp_details}
\begin{figure}
	\includegraphics[width=\linewidth]{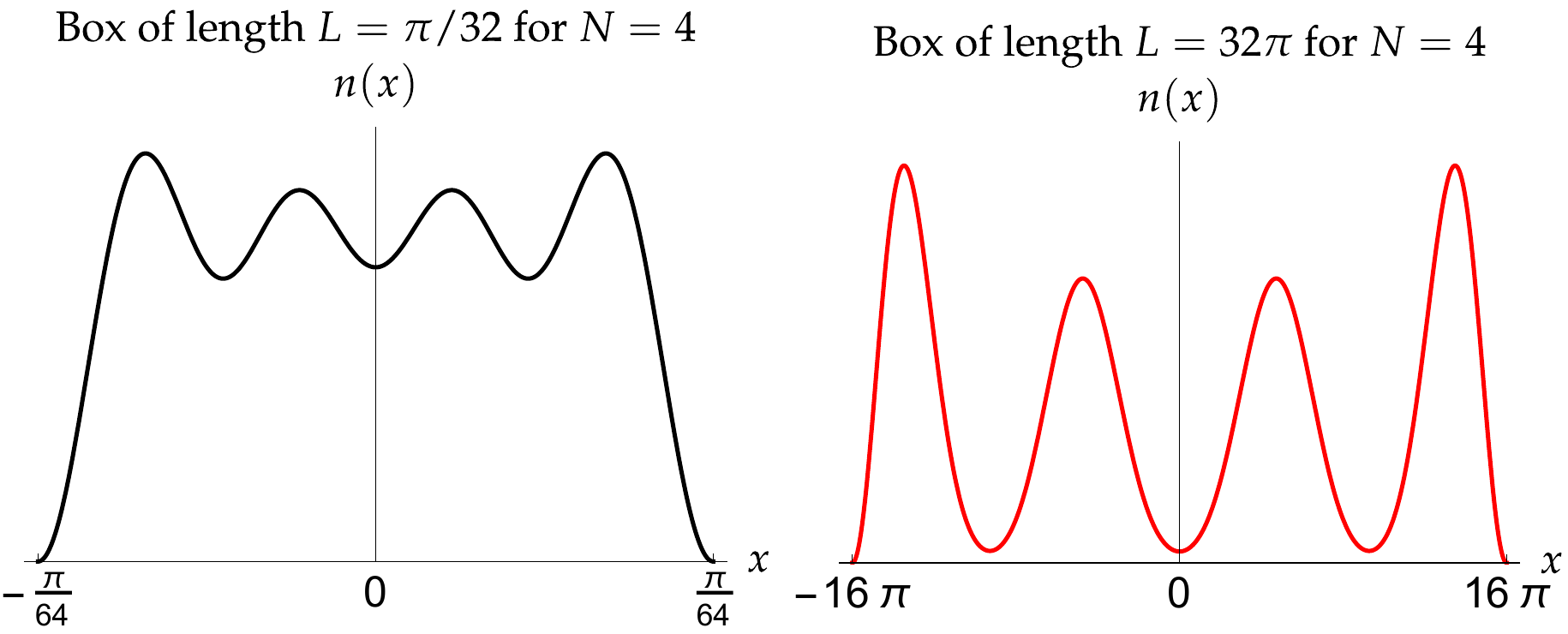}
	\caption{
	\titou{Ground-state one-electron density $\n{}{}(x)$ of 4-boxium (\ie, $N = 4$) for $L = \pi/32$ (left) and $L = 32\pi$ (right).
	In the weak correlation regime (small box length), the one-electron density is much more delocalized and uniform than in the strong correlation regime (large box length), where a Wigner crystal starts to appear. \cite{Rogers_2017,Rogers_2016}}
	\label{fig:rho}
	}
\end{figure}

Having defined the eLDA functional in the previous section [see Eq.~\eqref{eq:eLDA}], we now turn to its validation.
Our testing playground for the validation of the eLDA functional is the ubiquitous ``electrons in a box'' model where $\nEl$ electrons are confined in a 1D box of length $L$, a family of systems that we call $\nEl$-boxium in the following.
In particular, we investigate systems where $L$ ranges from $\pi/8$ to $8\pi$ and $2 \le \nEl \le 7$. 
These inhomogeneous systems have non-trivial electronic structure properties which can be tuned by varying the box length.
For small $L$, the system is weakly correlated, while strong correlation effects dominate in the large-$L$ regime. \cite{Rogers_2017,Rogers_2016}
\titou{The one-electron density in these two regimes of correlation is represented in Fig.~\ref{fig:rho}.}

We use as basis functions the (orthonormal) orbitals of the one-electron system, \ie,
\begin{equation}
	\AO{\mu}(x) = 
	\begin{cases}
		\sqrt{2/L} \cos(\mu \pi x/L),	&	\mu \text{ is odd,}
		\\
		\sqrt{2/L} \sin(\mu \pi x/L),	&	\mu \text{ is even,}
	\end{cases}
\end{equation} 
with $ \mu = 1,\ldots,\nBas$ and $\nBas = 30$ for all calculations.
The convergence threshold $\tau = \max{ \abs{ \bF{\bw} \bGam{\bw}
\bS - \bS \bGam{\bw} \bF{\bw}}}$ [see Eq.~\eqref{eq:commut_F_AO}] of the KS-DFT self-consistent calculation is set
to $10^{-5}$. 
In order to compute the various density-functional
integrals that cannot be performed in closed form, 
a 51-point Gauss-Legendre quadrature is employed.

In order to test the present eLDA functional we perform various sets of calculations.
To get reference excitation energies for both the single and double excitations, we compute full configuration interaction (FCI) energies with the Knowles-Handy FCI program described in Ref.~\onlinecite{Knowles_1989}.
For the single excitations, we also perform time-dependent LDA (TDLDA)
calculations [\ie, TDDFT with the LDA functional defined in Eq.~\eqref{eq:LDA}]. 
Its Tamm-Dancoff approximation version  (TDA-TDLDA) is also considered. \cite{Dreuw_2005}

Concerning the ensemble calculations, two sets of weight are tested: the zero-weight
(ground-state) limit where $\bw = (0,0)$ and the
equi-triensemble (or equal-weight state-averaged) limit where $\bw = (1/3,1/3)$.
\titou{Note that a zero-weight calculation does correspond to a ground-state KS calculation with $100\%$ exact exchange and LDA correlation.}

\section{Results and discussion}
\label{sec:res}

\begin{figure*}
	\includegraphics[width=\linewidth]{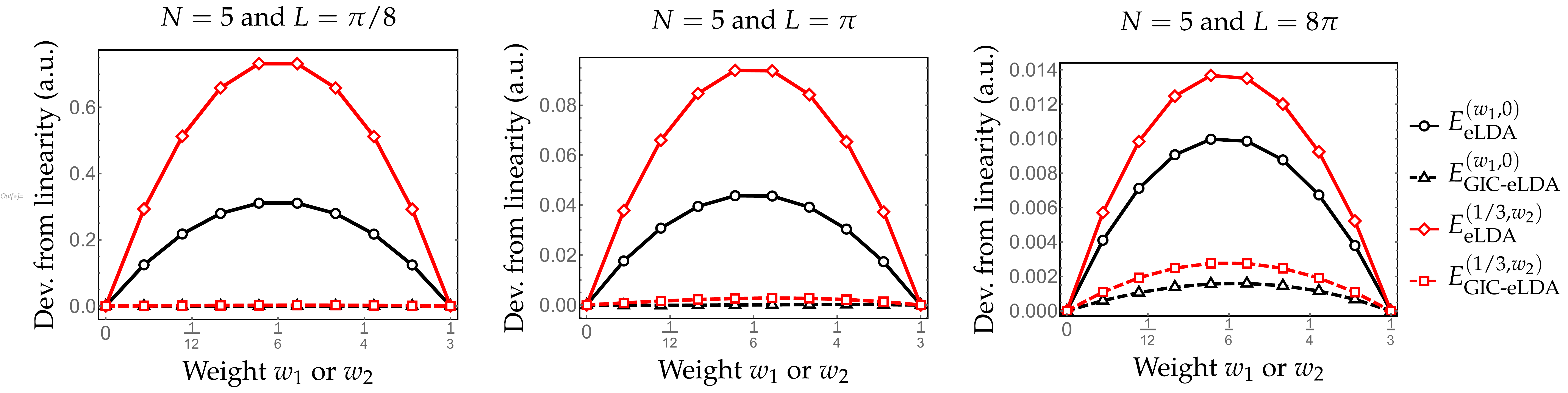}
	\caption{
	\label{fig:EvsW}
	Deviation from linearity of the weight-dependent KS-eLDA ensemble energy $\E{eLDA}{(\ew{1},\ew{2})}$ with (dashed lines) and without (solid lines) ghost-interaction correction (GIC) for 5-boxium (\ie, $\nEl = 5$) with a box of length $L = \pi/8$ (left), $L = \pi$ (center), and $L = 8\pi$ (right).
	}
\end{figure*}

First, we discuss the linearity of the computed (approximate)
ensemble energies.
To do so, we consider 5-boxium with box lengths of $L = \pi/8$, $L = \pi$, and $L = 8\pi$, which correspond (qualitatively at least) to the weak, intermediate, and strong correlation regimes, respectively.
The deviation from linearity of the three-state ensemble energy
$\E{}{(\ew{1},\ew{2})}$ (\ie, the deviation from the
linearly-interpolated ensemble energy) is represented
in Fig.~\ref{fig:EvsW} as a function of $\ew{1}$ or $\ew{2}$ while
fulfilling the restrictions on the ensemble weights to ensure the GOK
variational principle [\ie, $0 \le \ew{2} \le 1/3$ and $\ew{2} \le \ew{1} \le (1-\ew{2})/2$].
\manurev{More precisely, we follow a continuous path that connects
ground-state [$\bw=(0,0)$] and equiensemble [$\bw=(1/3,1/3)$]
calculations. For convenience, we use two connected paths. The first
one, for which $\ew{2}=0$ and $0\leq \ew{1}\leq 1/3$, relies on the 
biensemble while the second one is defined as follows:
$\ew{1}=1/3$ and $0\leq \ew{2}\leq 1/3$.}
To illustrate the magnitude of the ghost-interaction error, we report the KS-eLDA ensemble energy with and without GIC as explained above {[see Eqs.~\eqref{eq:Ew-GIC-eLDA} and \eqref{eq:Ew-eLDA}]}.
As one can see in Fig.~\ref{fig:EvsW}, without GIC, the
ensemble energy becomes less and less linear as $L$
gets larger, while the GIC reduces the curvature of the ensemble energy
drastically. 
It is important to note that, even though the GIC removes the explicit
quadratic Hx terms from the ensemble energy, a non-negligible curvature
remains in the GIC-eLDA ensemble energy when the electron
correlation is strong. \manurev{The latter ensemble energy is computed
as the weighted
sum of the individual KS-eLDA energies [see
Eq.~\eqref{eq:Ew-eLDA}]. Therefore, its 
curvature can only originate from the weight dependence of the
individual energies.
Note that such a dependence does not exist in the exact theory. Here,
the individual density-functional eLDA correlation energies exhibit an
explicit linear and quadratic dependence on the weights, as discussed
further in the next paragraph. Note also that the individual KS-eLDA energies
may gain an additional (implicit) dependence on the weights through the optimization of the
ensemble KS orbitals in the presence of ghost-interaction errors [see
Eqs.~\eqref{eq:min_with_HF_ener_fun} and \eqref{eq:WHF}].
}

\begin{figure*}
	\includegraphics[width=\linewidth]{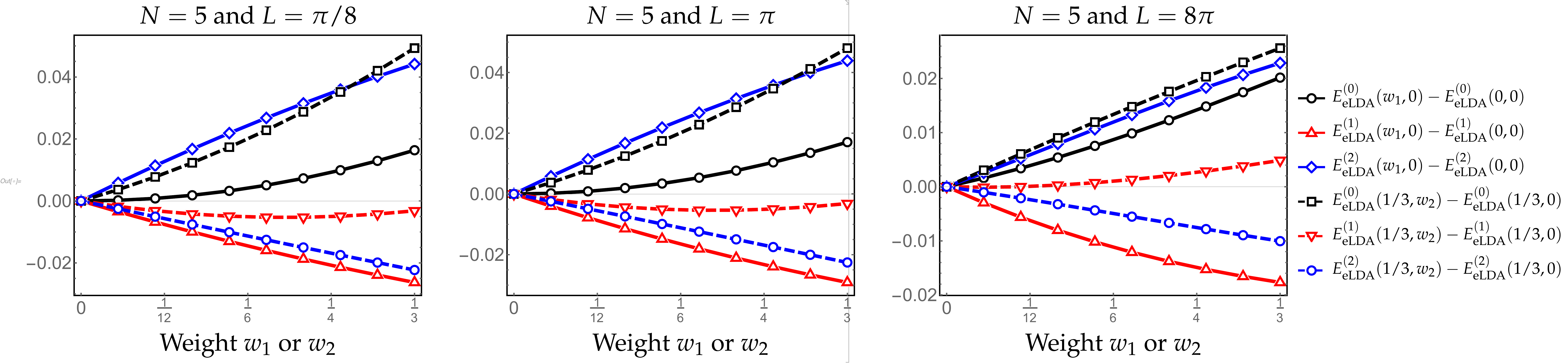}
	\caption{
	\label{fig:EIvsW}
	KS-eLDA individual energies, $\E{eLDA}{(0)}$ (black), $\E{eLDA}{(1)}$ (red), and $\E{eLDA}{(2)}$ (blue), as functions of the weights $\ew{1}$ (solid) and $\ew{2}$ (dashed) for 5-boxium (\ie, $\nEl = 5$) with a box of length $L = \pi/8$ (left), $L = \pi$ (center), and $L = 8\pi$ (right).}
\end{figure*}

Figure \ref{fig:EIvsW} reports the behavior of the three KS-eLDA individual energies as functions of the weights.
Unlike in the exact theory, we do not obtain
straight horizontal lines when plotting these
energies, which is in agreement with  
the curvature of the GIC-eLDA ensemble energy discussed previously. The variations in the ensemble weights are essentially linear or quadratic. 
\manurev{This can be rationalized as follows. As readily seen from
Eqs.~\eqref{eq:EI-eLDA} and \eqref{eq:ind_HF-like_ener}, the individual
HF-like energies do not depend explicitly on the weights, which means
that the above-mentioned variations originate from the eLDA correlation
functional [second and third terms on the right-hand side of
Eq.~\eqref{eq:EI-eLDA}]. If, for analysis purposes, we consider the
Taylor expansions around the uniform density regime in
Eqs.~\eqref{eq:Taylor_exp_ind_corr_ener_eLDA} and
\eqref{eq:Taylor_exp_DDisc_term}, contributions with an explicit weight
dependence still remain after summation. As both the ensemble density and
the ensemble correlation energy per particle vary linearly with the
weights $\bw$ [see Eqs.~\eqref{eq:ens1RDM},
\eqref{eq:ens_dens_from_ens_1RDM}, and
\eqref{eq:decomp_ens_correner_per_part}], the latter contributions will contain both linear and quadratic terms in
$\bw$, as evidenced by Eq.~\eqref{eq:Taylor_exp_DDisc_term} [see the second term on the right-hand
side].}\\ 
Interestingly, the
individual energies do not vary in the same way depending on the state
considered and the value of the weights. 
\titou{On one hand,} we see for example that, within the biensemble (\ie, $\ew{2}=0$), the energies of
the ground and \titou{second} excited-state increase with respect to the
first-excited-state weight $\ew{1}$, thus showing that, in this
case, we
``deteriorate'' these states by optimizing the orbitals for the
ensemble, rather than for each state separately. 
\titou{The singly excited state is, on the other hand, stabilized in the biensemble, which is reasonable as the weight associated with this state increases.
For the triensemble, as $\ew{2}$ increases, the energy of the ground state increases, while the energy of the first excited state remains stable with a slight increase at large $L$. 
The second excited state is obviously stabilized by the increase of its weight in the ensemble.
\manurev{
These are all very sensible observations.\\
Let us finally stress that the (well-known) poor performance of the
combined 100\% HF-exchange/LDA correlation scheme in 
ground-state [\ie, $\bw=(0,0)$] DFT, where the correlation energy is
overestimated, is substantially improved for the
ground state within the equiensemble [$\bw=(1/3,1/3)$]} (see the {\SI} for
further details).
This is a
remarkable and promising result. A similar improvement is observed for
the first excited state, at least in the weak correlation regime,
without deteriorating too much the second-excited-state energy.}

\begin{figure}
	\includegraphics[width=\linewidth]{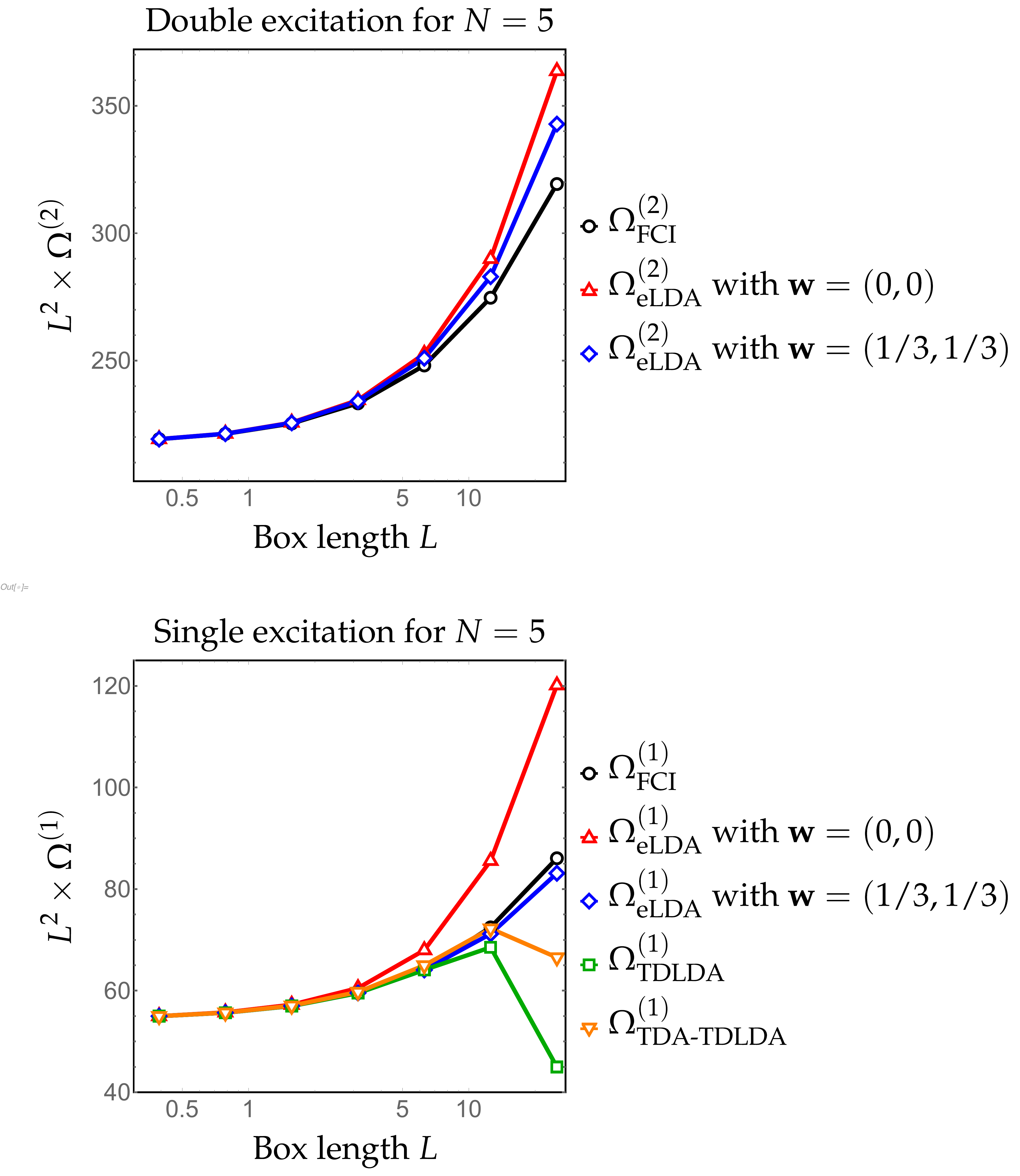}
	\caption{
	\label{fig:EvsL}
	Excitation energies (multiplied by $L^2$) associated with the single excitation $\Ex{}{(1)}$ (bottom) and double excitation $\Ex{}{(2)}$ (top) of 5-boxium for various methods and box lengths $L$.
	Graphs for additional values of $\nEl$ can be found as {\SI}.
	}
\end{figure}

Figure \ref{fig:EvsL} reports the excitation energies (multiplied by $L^2$) for various methods and box lengths in the case of 5-boxium (\ie, $\nEl = 5$).
Similar graphs are obtained for the other $\nEl$ values and they can be found in the {\SI} alongside the numerical data associated with each method.
For small $L$, the single and double excitations can be labeled as
``pure'', as revealed by a thorough analysis of the FCI wavefunctions.
In other words, each excitation is dominated by a sole, well-defined reference Slater determinant.
However, when the box gets larger (\ie, as $L$ increases), there is a strong mixing between the different excitation degrees.
In particular, the single and double excitations strongly mix, which makes their assignment as single or double excitations more disputable. \cite{Loos_2019}
This can be clearly evidenced by the weights of the different
configurations in the FCI wave function.

As shown in Fig.~\ref{fig:EvsL}, all methods provide accurate estimates of the excitation energies in the weak correlation regime (\ie, small $L$).
When the box gets larger, they start to deviate.
For the single excitation, TDLDA is extremely accurate up to $L = 2\pi$, but yields more significant errors at larger $L$ by underestimating the excitation energies.
TDA-TDLDA slightly corrects this trend thanks to error compensation. 
Concerning the eLDA functional, our results clearly evidence that the equiweight [\ie, $\bw = (1/3,1/3)$] excitation energies are much more accurate than the ones obtained in the zero-weight limit [\ie, $\bw = (0,0)$].
This is especially true, in the strong correlation regime, for the single excitation
which is significantly improved by using equal weights.
The effect on the double excitation is less pronounced.
Overall, one clearly sees that, with 
equal weights, KS-eLDA yields accurate excitation energies for both single and double excitations.
This conclusion is verified for smaller and larger numbers of electrons
(see {\SI}).
\titou{Except for the two-electron system where we observe cases of underestimation, eLDA usually overestimates double excitations, as evidenced by the numerical data gathered in the {\SI}.}

\begin{figure*}
	\includegraphics[width=\linewidth]{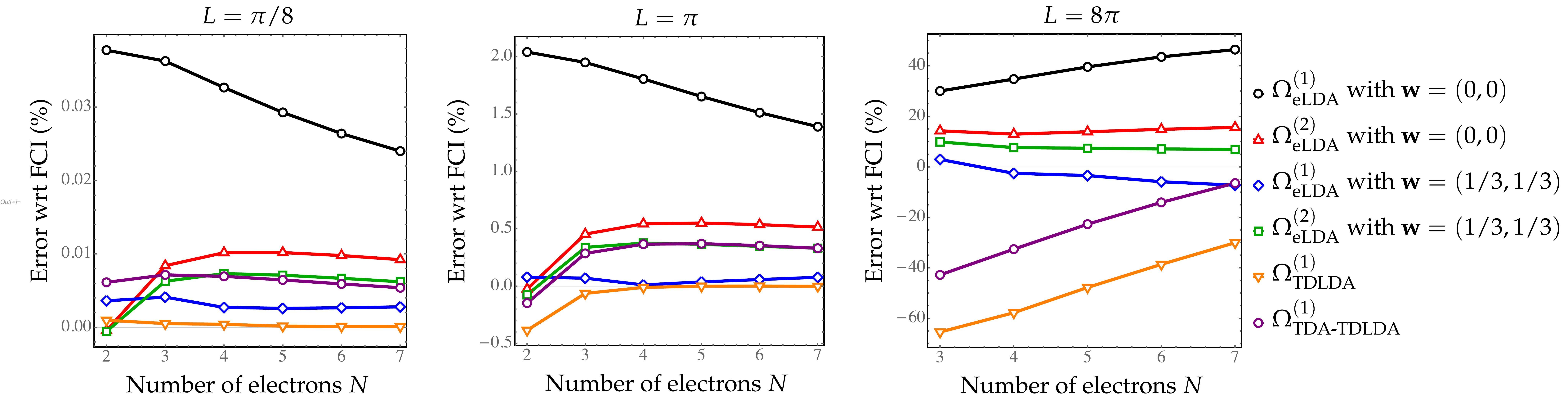}
	\caption{
	\label{fig:EvsN}
	Error with respect to FCI in single and double excitation energies for $\nEl$-boxium for various methods and electron numbers $\nEl$ at $L=\pi/8$ (left), $L=\pi$ (center), and $L=8\pi$ (right).
	}
\end{figure*}

For the same set of methods, Fig.~\ref{fig:EvsN} reports the error (in \%) in excitation energies (as compared to FCI) as a function of $\nEl$ for three values of $L$ ($\pi/8$, $\pi$, and $8\pi$).
We draw similar conclusions as above: irrespectively of the number of
electrons, the eLDA functional with equal
weights is able to accurately model single and double excitations, with
a very significant improvement brought by the 
equiensemble KS-eLDA orbitals as compared to their zero-weight
(\ie, conventional ground-state) analogs.
As a rule of thumb, in the weak and intermediate correlation regimes, we
see that the single
excitation obtained from equiensemble KS-eLDA is of
the same quality as the one obtained in the linear response formalism
(such as TDLDA). On the other hand, the double
excitation energy only deviates
from the FCI value by a few tenth of percent.
Moreover, we note that, in the strong correlation regime
(right graph of Fig.~\ref{fig:EvsN}), the single excitation
energy obtained at the equiensemble KS-eLDA level remains in good
agreement with FCI and is much more accurate than the TDLDA and TDA-TDLDA excitation energies which can deviate by up to $60 \%$.
This also applies to the double excitation, the discrepancy
between FCI and equiensemble KS-eLDA remaining of the order of a few percents in the strong correlation regime.
These observations nicely illustrate the robustness of the
GOK-DFT scheme in any correlation regime for both single and double excitations.
This is definitely a very pleasing outcome, which additionally shows
that, even though we have designed the eLDA functional based on a
two-electron model system, the present methodology is applicable to any
1D electronic system, \ie, a system that has more than two
electrons. 

\begin{figure*}
	\includegraphics[width=\linewidth]{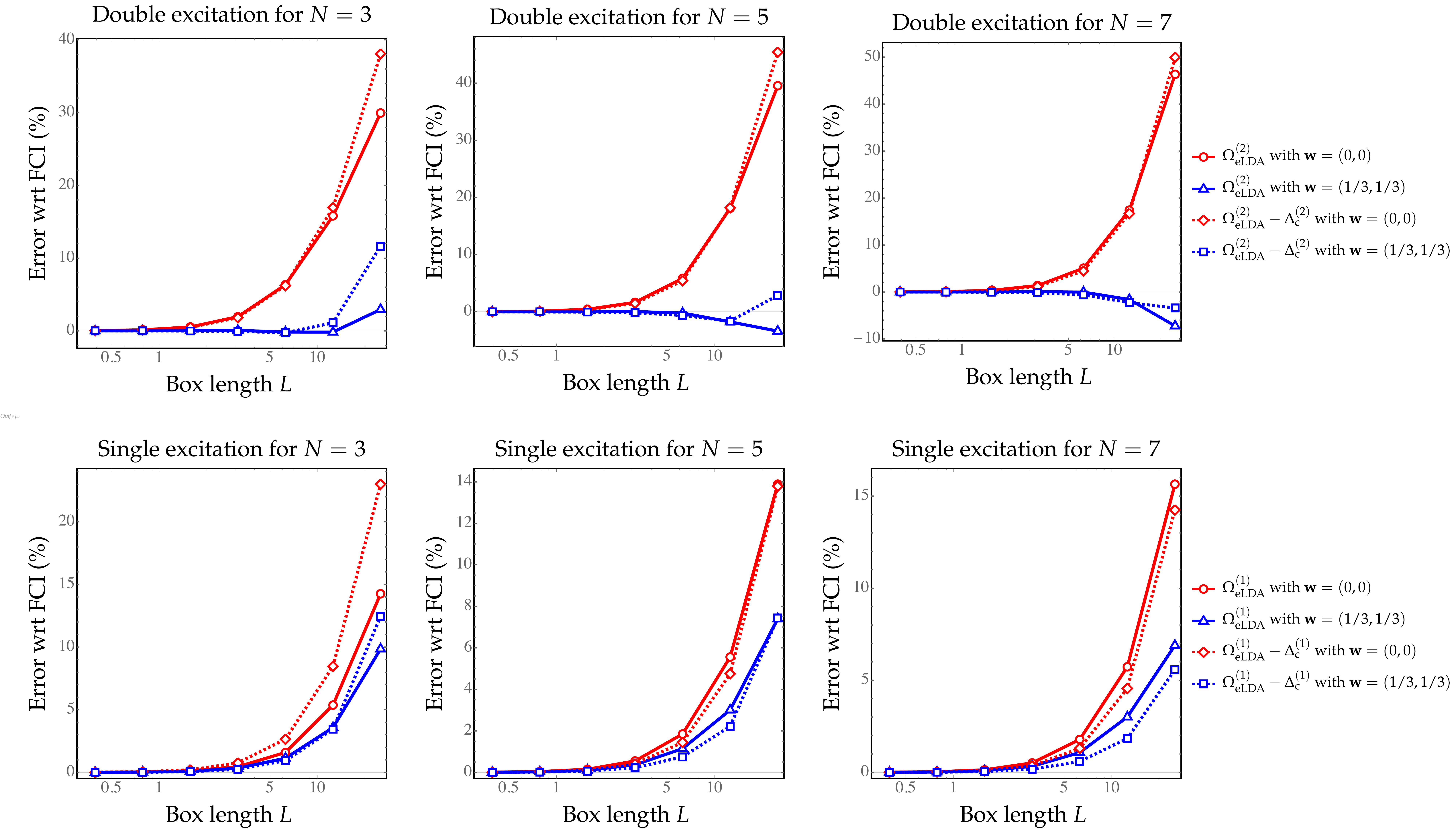}
	\caption{
	\label{fig:EvsL_DD}
	Error with respect to FCI (in \%) associated with the single excitation $\Ex{}{(1)}$ (bottom) and double excitation $\Ex{}{(2)}$ (top) as a function of the box length $L$ for 3-boxium (left), 5-boxium (center), and 7-boxium (right) at the KS-eLDA level with and without the contribution of the ensemble correlation derivative $\DD{c}{(I)}$.
	Zero-weight (\ie, $\ew{1} = \ew{2} = 0$, red lines) and equiweight (\ie, $\ew{1} = \ew{2} = 1/3$, blue lines) calculations are reported. 
	}
\end{figure*}

It is also interesting to investigate the influence of the
correlation ensemble derivative contribution $\DD{c}{(I)}$
to the $I$th excitation energy [see Eq.~\eqref{eq:DD-eLDA}].
In our case, both single ($I=1$) and double ($I=2$) excitations are considered.
To do so, we have reported in Fig.~\ref{fig:EvsL_DD}, for $\nEl = 3$, $5$, and $7$, the error percentage (with respect to FCI) as a function of the box length $L$
on the excitation energies obtained at the KS-eLDA level with and without $\DD{c}{(I)}$ [\ie, the last term in Eq.~\eqref{eq:Om-eLDA}].
We first stress that although for $\nEl=3$ both single and double excitation energies are 
systematically improved (as the strength of electron correlation
increases) when
taking into account  
the correlation ensemble derivative, this is not
always the case for larger numbers of electrons.
For 3-boxium, in the zero-weight limit, the correlation ensemble derivative is
significantly larger for the single
excitation as compared to the double excitation; the reverse is observed in the equal-weight triensemble
case.
However, for 5- and 7-boxium, it hardly
influences the double excitation (except when the correlation is strong), and slightly deteriorates the single excitation in the intermediate and strong correlation regimes.
This non-systematic behavior in terms of the number of electrons might
be a consequence of how we constructed eLDA.
Indeed, as mentioned in Sec.~\ref{sec:eDFA}, the weight dependence of
the eLDA functional is based on a \textit{two-electron} finite uniform electron gas.
Incorporating a $\nEl$-dependence in the functional through the
curvature of the Fermi hole, in the spirit of Ref.~\onlinecite{Loos_2017a}, would be
valuable in this respect. This is left for future work.
Interestingly, for the single excitation in 3-boxium, the magnitude of the correlation ensemble 
derivative is substantially reduced when switching from a zero-weight to
an equal-weight calculation, while giving similar excitation energies,
even in the strongly correlated regime. A possible interpretation is
that, at least for the single excitation, equiensemble orbitals partially remove the burden
of modelling properly the correlation ensemble derivative. 
This conclusion does not hold for larger
numbers of electrons ($N=5$ or $7$), possibly because eLDA extracts density-functional correlation ensemble
derivatives from a two-electron uniform electron gas, as mentioned previously.
For the double excitation, the ensemble derivative remains important, even in
the equiensemble case. 
To summarize, the equiensemble calculation 
is always more accurate than a zero-weight
(\ie, a conventional ground-state DFT) one, with or without including the ensemble 
derivative correction. Note that the second term on the right-hand side
of
Eq.~\eqref{eq:Om-eLDA}, which involves the weight-dependent correlation
potential and the density difference between ground and excited states,
has a negligible effect on the excitation energies (results not
shown).

\begin{figure}
	\includegraphics[width=\linewidth]{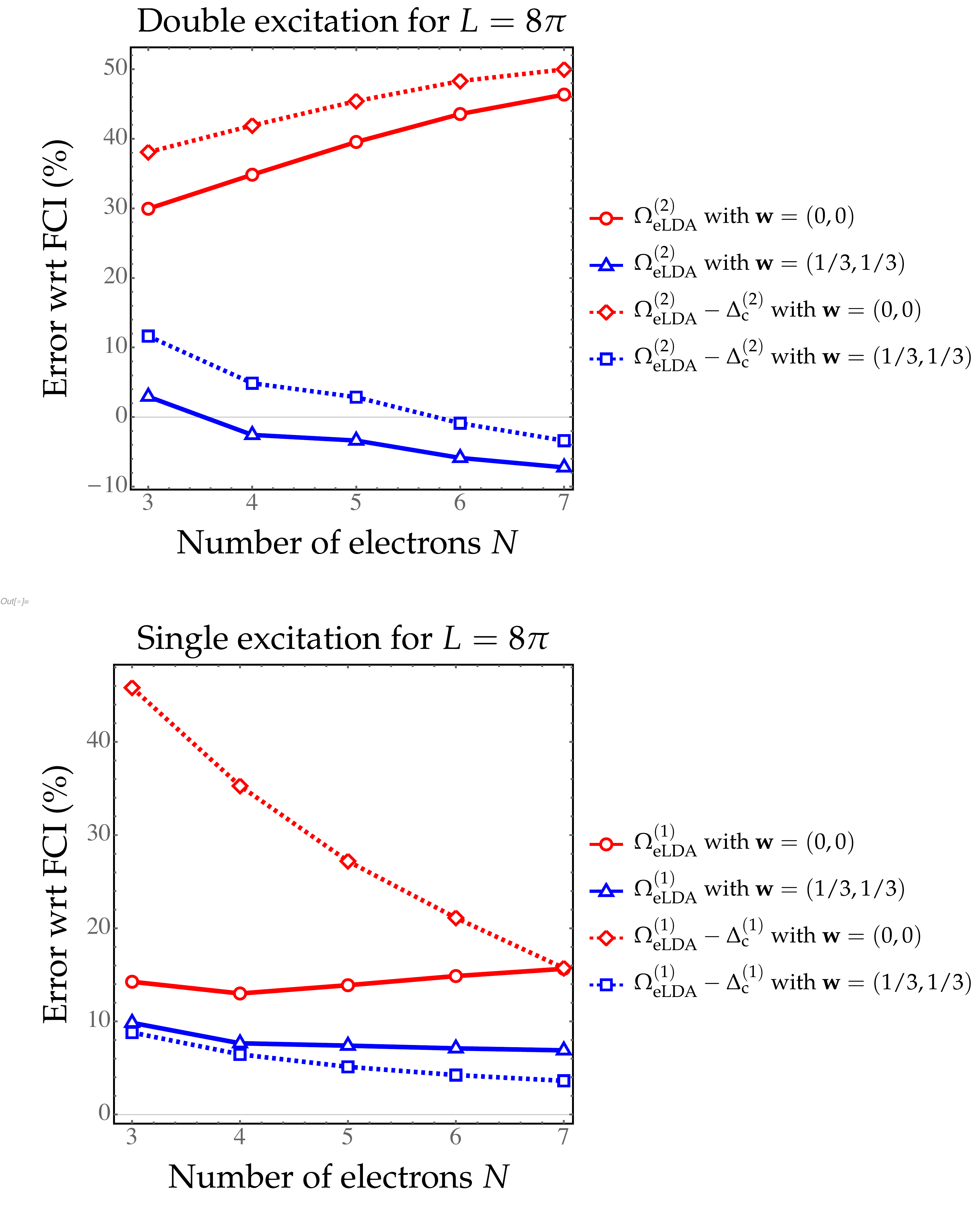}
	\caption{
	\label{fig:EvsN_DD}
	Error with respect to FCI in single and double excitation energies for $\nEl$-boxium (with a box length of $L=8\pi$) as a function of the number of electrons $\nEl$ at the KS-eLDA level with and without the contribution of the ensemble correlation derivative $\DD{c}{(I)}$. 
	Zero-weight (\ie, $\ew{1} = \ew{2} = 0$, red lines) and equiweight (\ie, $\ew{1} = \ew{2} = 1/3$, blue lines) calculations are reported. 
	}
\end{figure}

Finally, in Fig.~\ref{fig:EvsN_DD}, we report the same quantities as a function of the electron number for a box of length $8\pi$ (\ie, in the strong correlation regime). 
The difference between the solid and dashed curves
undoubtedly show that the
correlation ensemble derivative has a rather significant impact on the double
excitation (around $10\%$) with a slight tendency of worsening the excitation energies
in the case of equal weights, as the number of electrons
increases. It has a rather large influence (which decreases with the
number of electrons) on the single
excitation energies obtained in the zero-weight limit, showing once
again that the usage of equal weights has the benefit of significantly reducing the magnitude of the correlation ensemble derivative.

\section{Concluding remarks}
\label{sec:conclusion}

A local and ensemble-weight-dependent correlation density-functional approximation
(eLDA) has been constructed in the context of GOK-DFT for spin-polarized
triensembles in
1D. The approach is general and can be extended to real
(three-dimensional)
systems~\cite{Loos_2009,Loos_2009c,Loos_2010,Loos_2010d,Loos_2017a} 
and larger ensembles in order to  
model excited states in molecules and solids. Work is currently in
progress in this direction.

Unlike any standard functional, eLDA incorporates derivative
discontinuities through its weight dependence. The latter originates
from the finite uniform electron gas on which eLDA is
(partially) based. The KS-eLDA scheme, where exact individual
exchange energies are
combined with the eLDA correlation functional	, delivers accurate excitation energies for both
single and double excitations, especially when an equiensemble is used.
In the latter case, the same weights are assigned to each state belonging to the ensemble.
The improvement on the excitation energies brought by the KS-eLDA scheme is particularly impressive in the strong correlation regime where usual methods, such as TDLDA, fail.
We have observed that, although the correlation ensemble derivative has a
non-negligible effect on the excitation energies (especially for the
single excitations), its magnitude can be significantly reduced by
performing equiweight calculations instead of zero-weight
calculations.

Let us finally stress that the present methodology can be extended to other types of ensembles like, for example, the
$\nEl$-centered ones, \cite{Senjean_2018,Senjean_2020} thus allowing for the design of a LDA-type functional for the
calculation of ionization potentials, electron affinities, and
fundamental gaps. 
Like in the present
eLDA, such a functional would incorporate the infamous derivative
discontinuity contribution to the fundamental gap through its explicit weight
dependence. We hope to report on this in the near future.

\section*{Supplementary material}
See {\SI} for the additional details about the construction of the functionals, raw data and additional graphs.

\section*{Data availability statement}
The data that supports the findings of this study are available within the article [and its supplementary material].

\begin{acknowledgements}
The authors thank Bruno Senjean and Clotilde Marut for stimulating discussions.
This work has been supported through the EUR grant NanoX ANR-17-EURE-0009 in the framework of the \textit{``Programme des Investissements d'Avenir''.} 
\end{acknowledgements}

%
\end{document}